\documentclass{iau} 
\usepackage{graphicx}

\title[A two-component description of mass-segregated, anisotropic globular clusters] 
{A simple two-component description of mass segregation for anisotropic globular clusters}

\author[S. Torniamenti, G. Bertin \& P. Bianchini]   
{S. Torniamenti$^1$
 \and G. Bertin$^1$ \and P. Bianchini$^2$}

\affiliation{$^1$Universit\`{a} degli Studi di Milano, Dipartimento di Fisica, via Celoria 16, 20133 Milano, Italy \\  email: {\tt stefano.torniamenti@studenti.unimi.it} \\ email: {\tt giuseppe.bertin@unimi.it} \\[\affilskip]
$^2$Observatoire Astronomique de Strasbourg, 11 Rue de l'Universit\'{e}, 67000 Strasbourg, France \\email: {\tt paolo.bianchini@astro.unistra.fr}}

\pubyear{2019}
\volume{351}  
\jname{Star Clusters: From the Milky Way to the Early Universe}
\editors{A. Bragaglia, M.B. Davies, A. Sills \& E. Vesperini, eds.}
\begin{document}

\maketitle

\begin{abstract}
As a result of the slow action of two-body encounters, globular clusters develop mass segregation and attain a condition of only partial energy equipartition even in their central, most relaxed regions. Realistic numerical simulations show that, during the process, a radially-biased anisotropy profile slowly builds up, mimicking that resulting from incomplete violent relaxation. Commonly used dynamical models, such as the one-component King models, cannot describe these properties. Here we show that simple two-component models based on a distribution function originally conceived to describe elliptical galaxies, recently truncated and adapted to the context of globular clusters, can describe in detail what is observed in complex and realistic numerical simulations.
\end{abstract}

\firstsection 
\section{Introduction}

Improved numerical simulations and observations (especially, measurements of proper motions) are giving new insights into the processes of mass segregation and energy equipartition in globular clusters. Numerical investigations have revealed that these systems can attain a condition of only partial energy equipartition even in their central, most relaxed regions (\cite[Trenti \& van der Marel 2013]{trenti13}). Evidence for partial energy equipartition (e.g., see \cite[Libralato et al. 2018]{libralato18}) and a certain degree of mass segregation (e.g., see \cite[van der Marel \& Anderson 2010]{vandermarel10}; \cite[Di Cecco et al. 2013]{dicecco13}) has also been found observationally.

A rather unexpected phenomenon brought out by numerical simulations is the fact that, during the slow relaxation process, a radially-biased pressure anisotropy profile slowly builds up, even from initially isotropic conditions (\cite[Tiongco et al. 2016]{tiongco16}; \cite[Bianchini et al. 2017]{bianchini17}), mimicking the anisotropy profiles resulting from incomplete violent relaxation, a process known to be relevant to the formation of elliptical galaxies. The presence of radially-biased pressure anisotropy has also been noted in recent observations (Watkins et al. 2015; Bellini et al. 2017). We recall that the most successful and  widely used models of globular clusters, the King (1966) models, are isotropic.

In stellar dynamics, to take into account energy equipartition and mass segregation we must consider multi-component models (e.g., see the 10-component models by \cite[Da Costa \& Freeman 1976]{dacosta76}). Here we address the ambitious goal of reducing the representation to the simplest possible case of two-component models, with the additional aim of describing also the relevant pressure anisotropy profile. We thus start from a distribution function originally conceived to describe elliptical galaxies (the $f^{(\nu)}$ function, \cite[Bertin \& Stiavelli 1993]{bertin93}), recently truncated and adapted to the context of globular clusters (\cite[de Vita et al. 2016]{devita16}). We then test the quality of our simple, self-consistent, physically-justified, two-component models by applying them to a set of realistic simulated states. A full description of this investigation is given in a separate extended article (\cite[Torniamenti et al. 2019]{torniamenti19}).

\section{Models and simulations}
\subsection{Models}
We consider the family of two-component self-consistent models constructed from the distribution function (\cite[de Vita et al. 2016]{devita16}):
\begin{equation}
f_{T,i}^{(\nu)} (E,J)=A_i \exp\left[-a_i (E-E_i)-d_i \frac{J}{|E-E_t|^{3/4}}\right]
\end{equation}
for $E<E_t$ and $f_{T,i}^{(\nu)} (E,J)=0$ otherwise. Here $E$ is the specific energy of a single star and $J$ represents the magnitude of the single star specific angular momentum. The quantities $A_i$, $a_i$, and $d_i$ are positive constants referring to the i-th component and $E_t$ is the truncation energy. 

The index $i$ labels the two species, which are set to be light stars of mass $m_1$ and heavy stars of mass $m_2 > m_1$. As described in detail by \cite[de Vita et al. (2016)]{devita16} and \cite[Torniamenti et al. (2019)]{torniamenti19}, by means of physical arguments the number of dimensionless parameters characterizing this family of two-component models can be reduced to two ($\Psi$,$\gamma$), the depth of the central potential well and the truncation parameter respectively.

\subsection{Simulations}

We consider the set of Monte Carlo cluster simulations developed and performed by \cite[Downing et al. (2010)]{downing10} that include both stellar and dynamical evolution. The detailed description of the initial conditions of the simulations are reported in \cite[Downing et al. (2010)]{downing10} and \cite[Bianchini et al. (2016)]{bianchini16}. In our study, eight snapshots of the simulations with  different levels of relaxation are studied as simulated states.

We identify the light component of the models with the main sequence stars (with mass $m_{1}$ equal to their mean mass) and combine giant stars, dark remnants (white dwarfs, neutron stars, and black holes), and binaries into the heavy component (with $m_{2}$ equal to the mean star mass of these three classes). A condition of only partial energy equipartition, quantified by the parameter $\eta$ defined in \cite{trenti13}, is established between the two species in all the simulated states, with $\eta\leq0.27$. We fix the value of $\eta$ in our fitting procedure to that of the simulated states. 

\begin{figure}[h]
	\centering
	\includegraphics[height=5 cm, width=9cm]{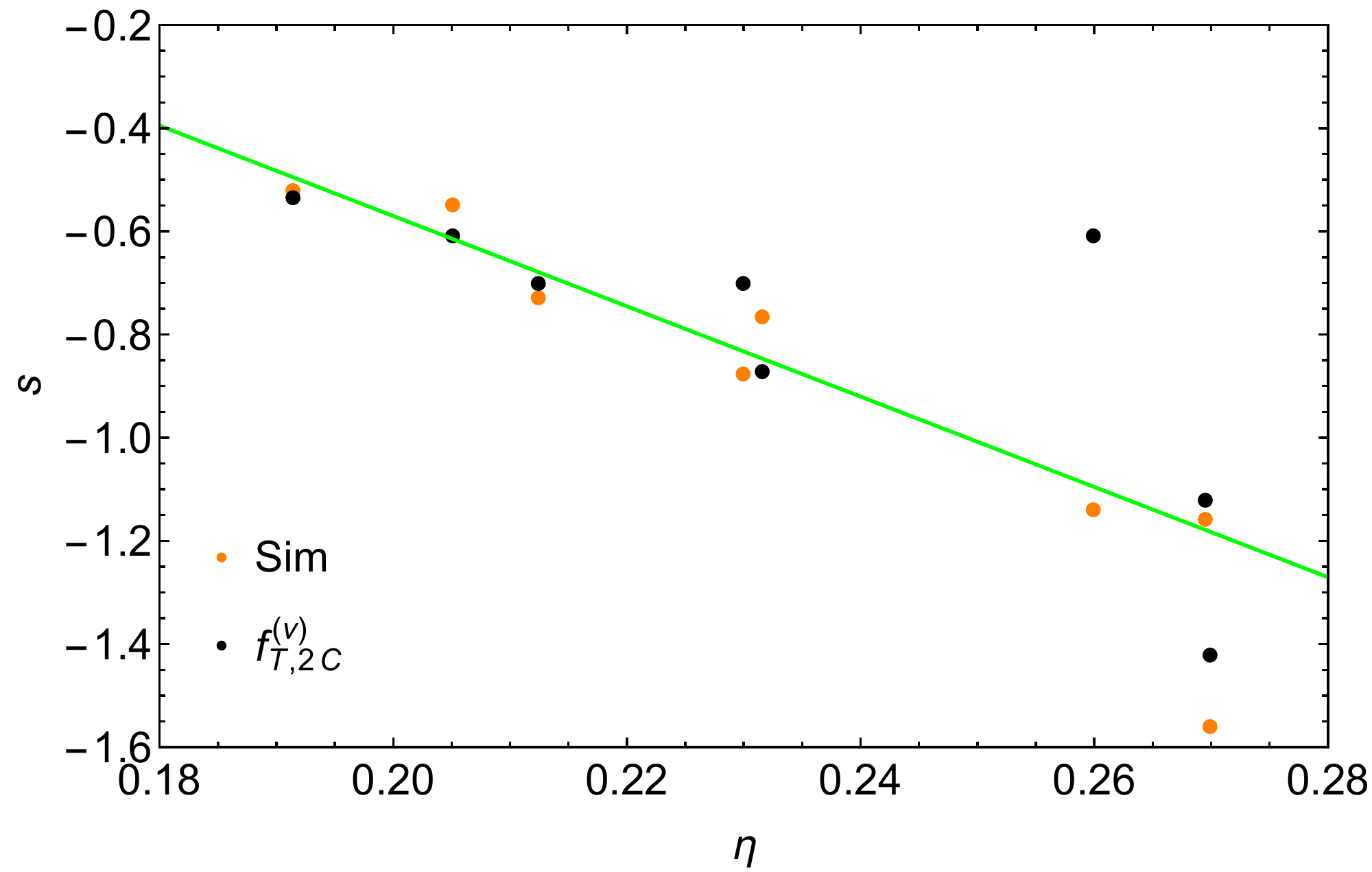}  
	\caption{Relation between the degree of mass segregation \textit{s} and the equipartition parameter $\eta$ for the simulated states (light, orange): a linear relation is found (light, green). Black dots show the corresponding points for the best-fit $f_T^{(\nu)}$ models.}
	\label{slope_eta_relation}%
\end{figure}

The degree of mass segregation is quantified by \textit{s}, that is the slope of the half-mass radius as a function of star mass. Figure \ref{slope_eta_relation} illustrates the relation between the degree of mass segregation \textit{s} and the equipartition parameter $\eta$ (orange dots), compared to that given by the best-fit two-component $f_T^{(\nu)}$ models (black dots, see Sect. \ref{fitting}). A linear relation is found between the two parameters.

\section{Complex simulated states fitted by simple two-component models} \label{fitting}

\begin{figure*}[ht!]
	\centering
	\begin{minipage}{0.49\textwidth}
		\centering
		\includegraphics[width=0.98\textwidth]{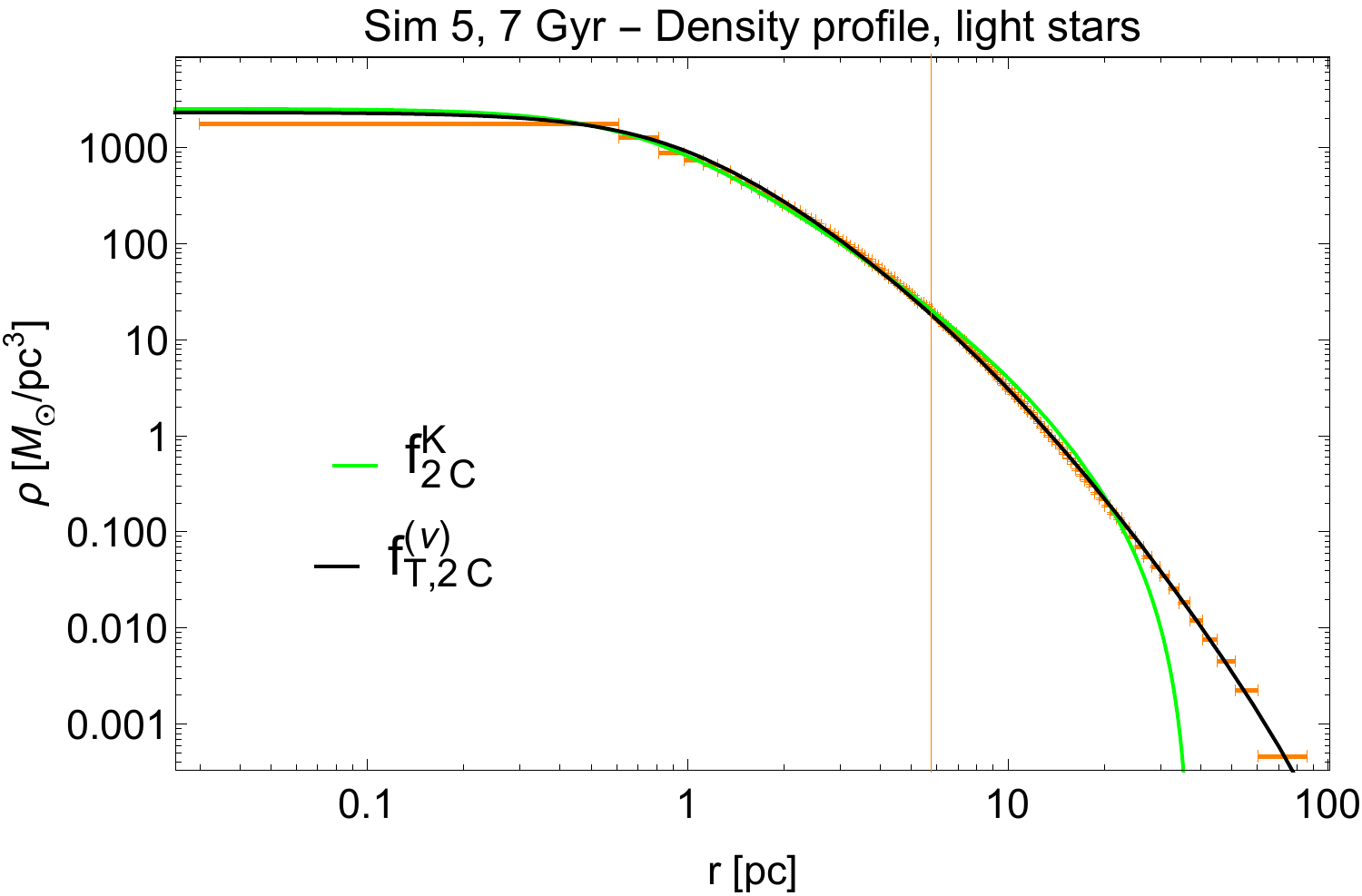}  
	\end{minipage}\hfill
	\begin{minipage}{0.49\textwidth}
		\centering
		\includegraphics[width=0.98\textwidth]{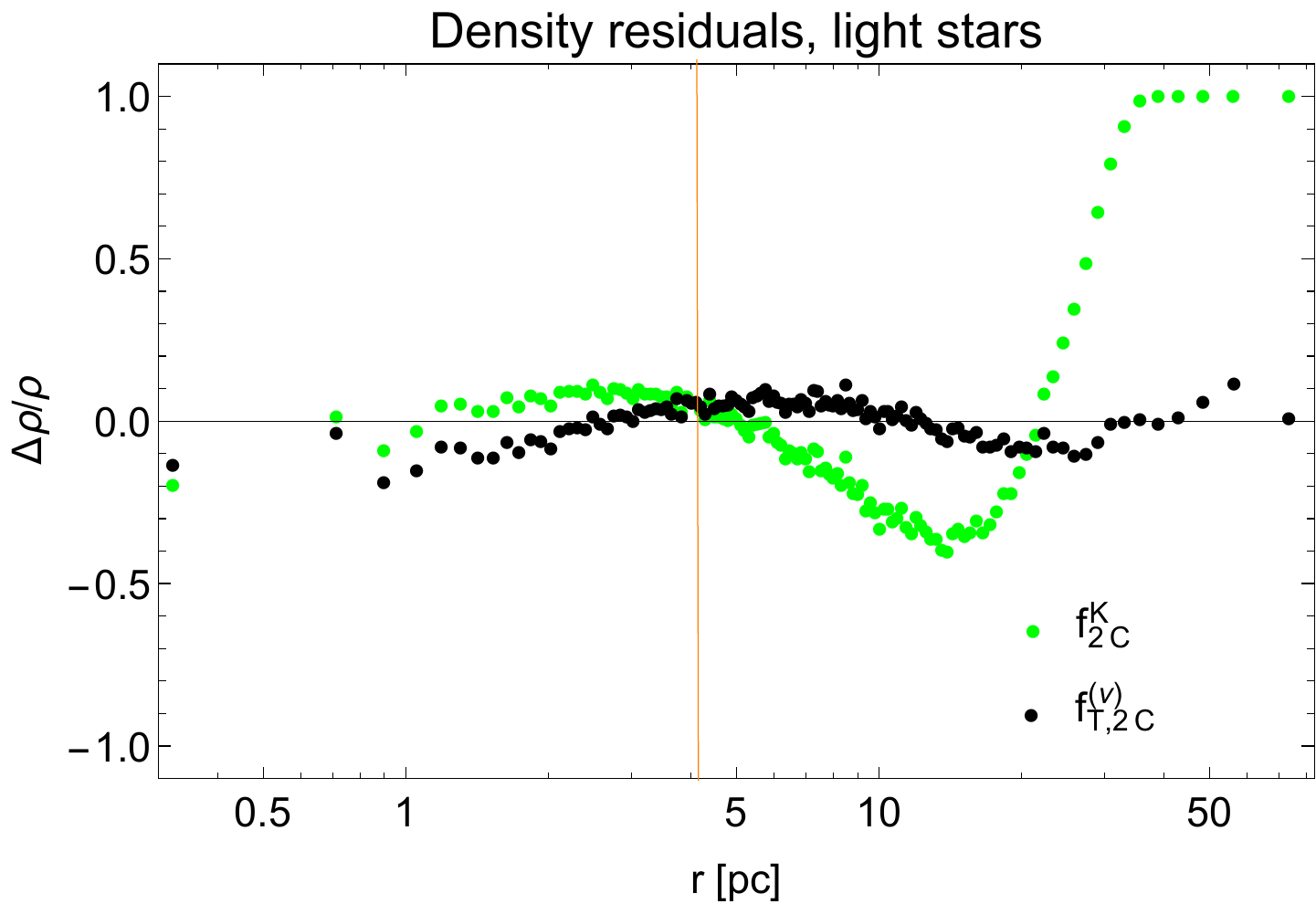} 
	\end{minipage}

	\begin{minipage}{0.49\textwidth}
		\centering
		\includegraphics[width=0.98\textwidth]{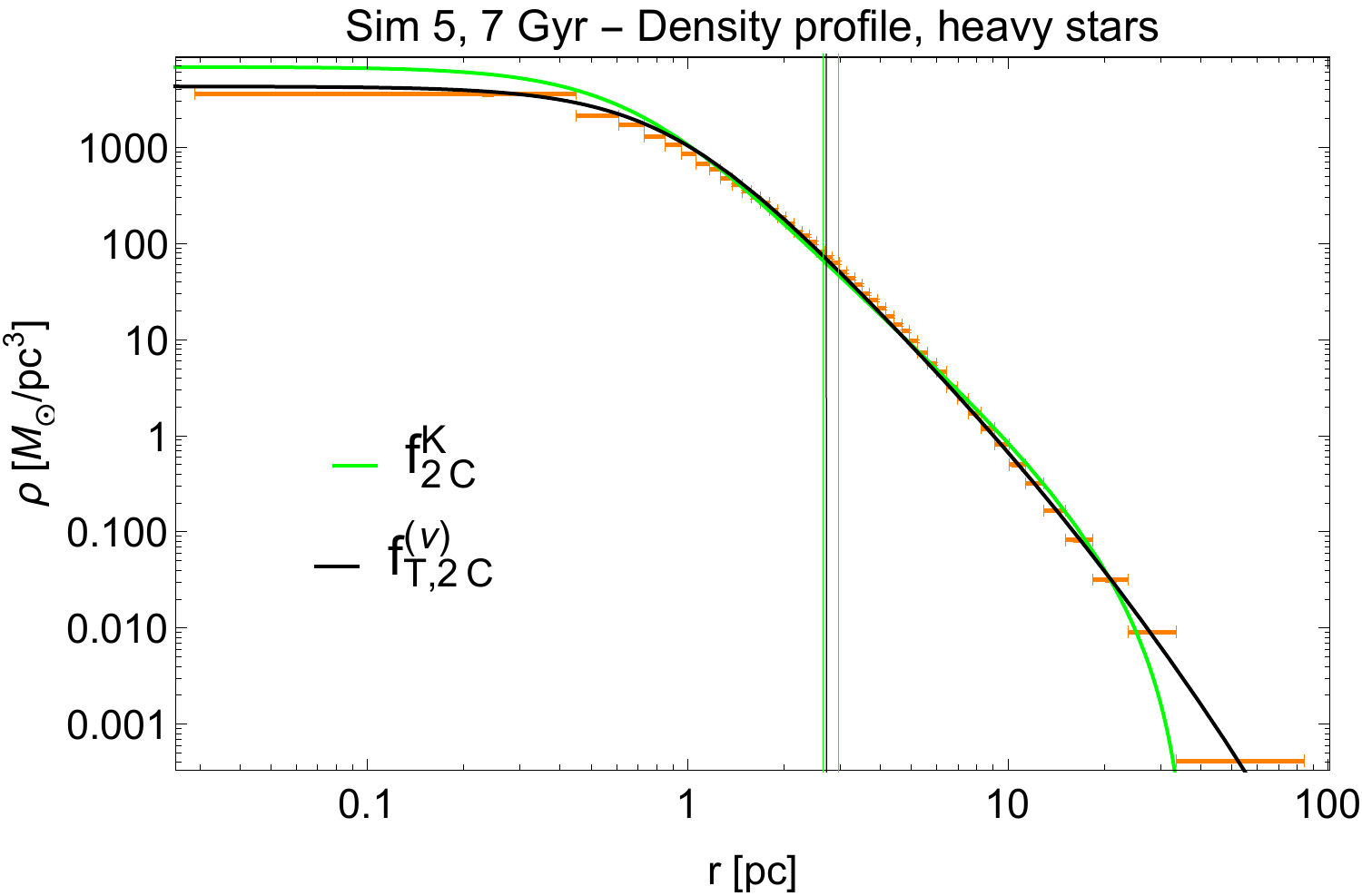} 
	\end{minipage} \hfill
	\begin{minipage}{0.49\textwidth}
		\centering
		\includegraphics[width=0.98\textwidth]{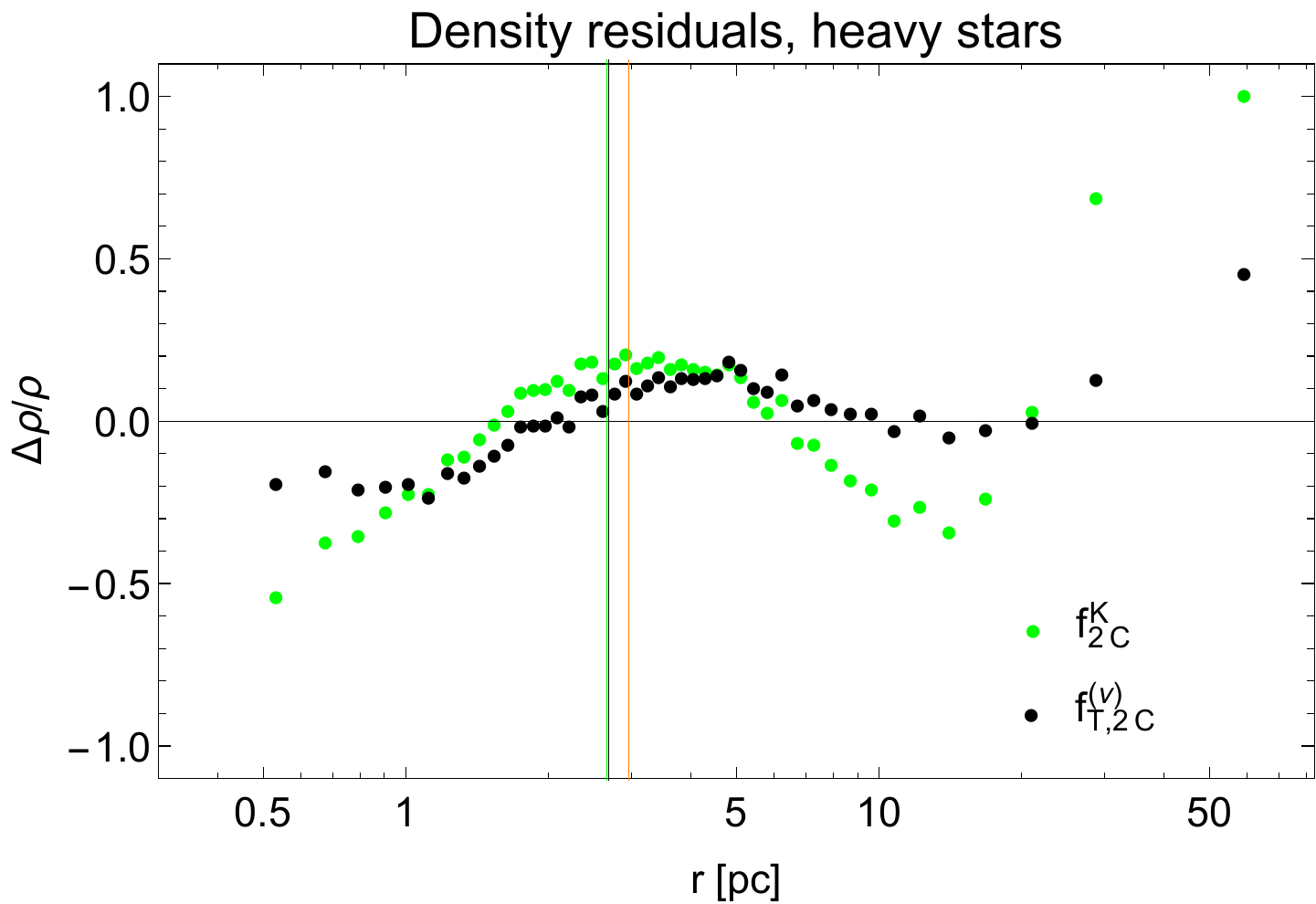} 
	\end{minipage} \hfill

	\begin{minipage}{0.49\textwidth}
		\centering
		\includegraphics[width=0.98\textwidth]{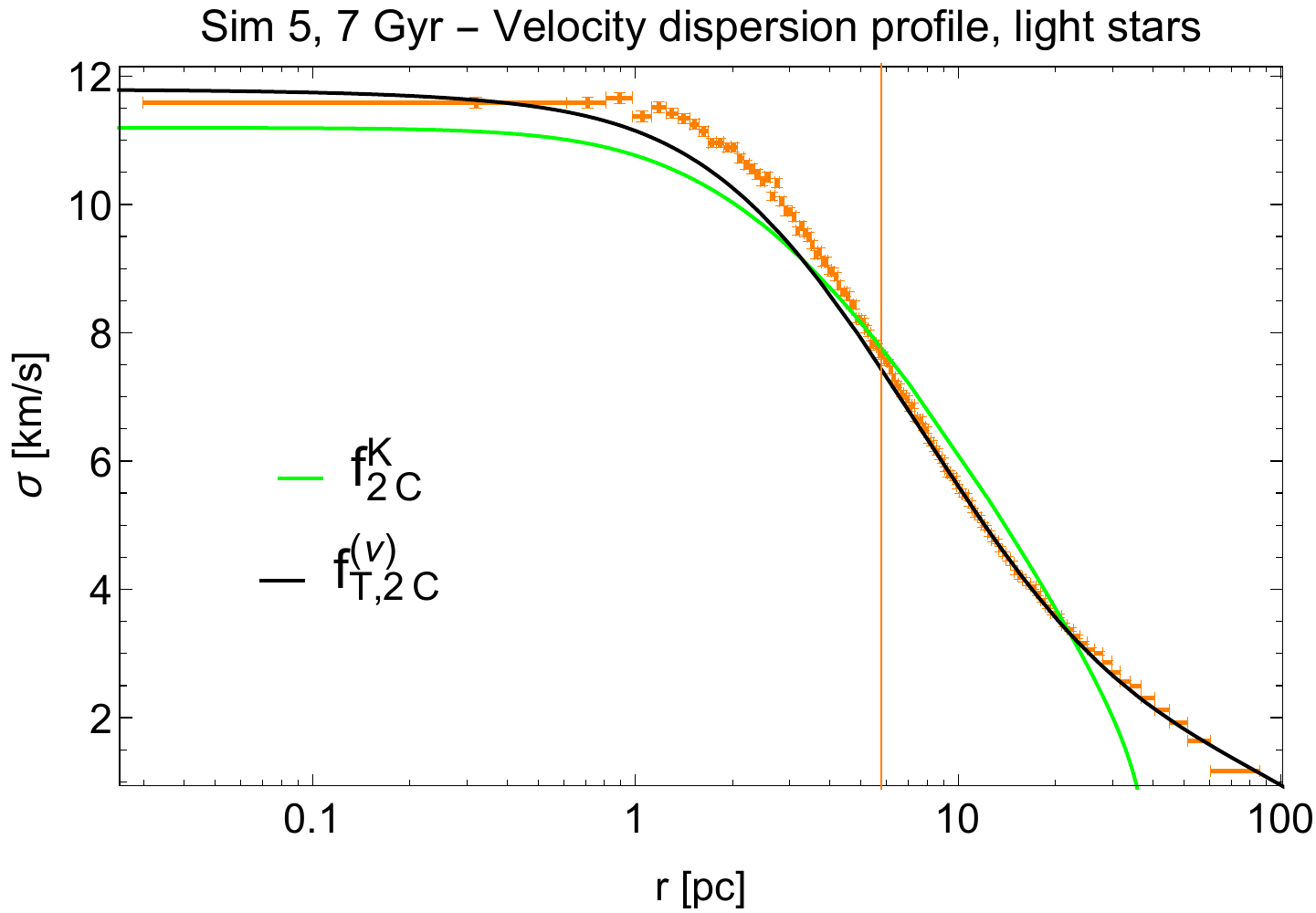}  
	\end{minipage}\hfill
	\begin{minipage}{0.49\textwidth}
		\centering
		\includegraphics[width=0.98\textwidth]{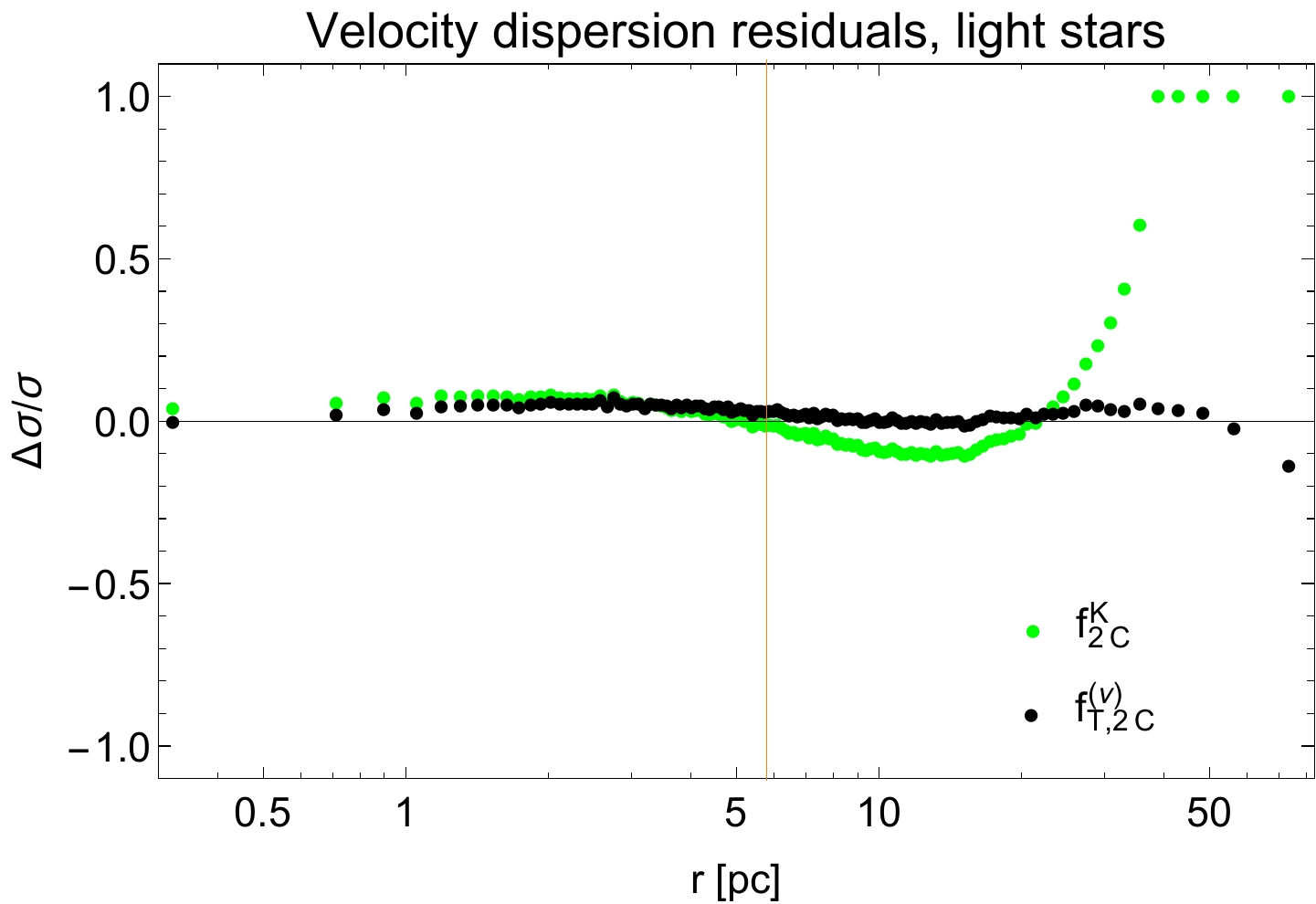} 
	\end{minipage} 
	
	\begin{minipage}{0.49\textwidth}
		\centering
		\includegraphics[width=0.98\textwidth]{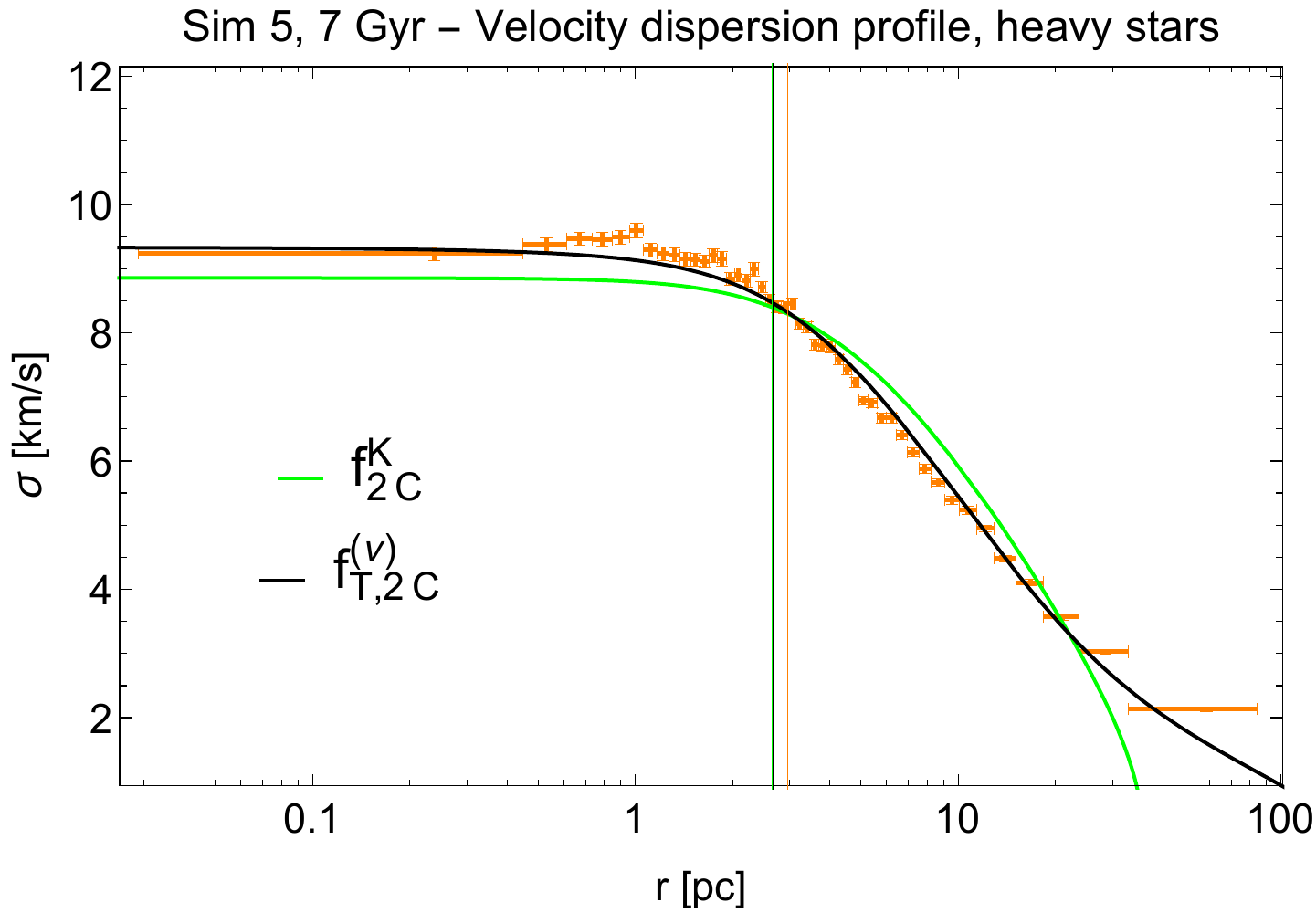} 
	\end{minipage} \hfill
	\begin{minipage}{0.49\textwidth}
		\centering
		\includegraphics[width=0.98\textwidth]{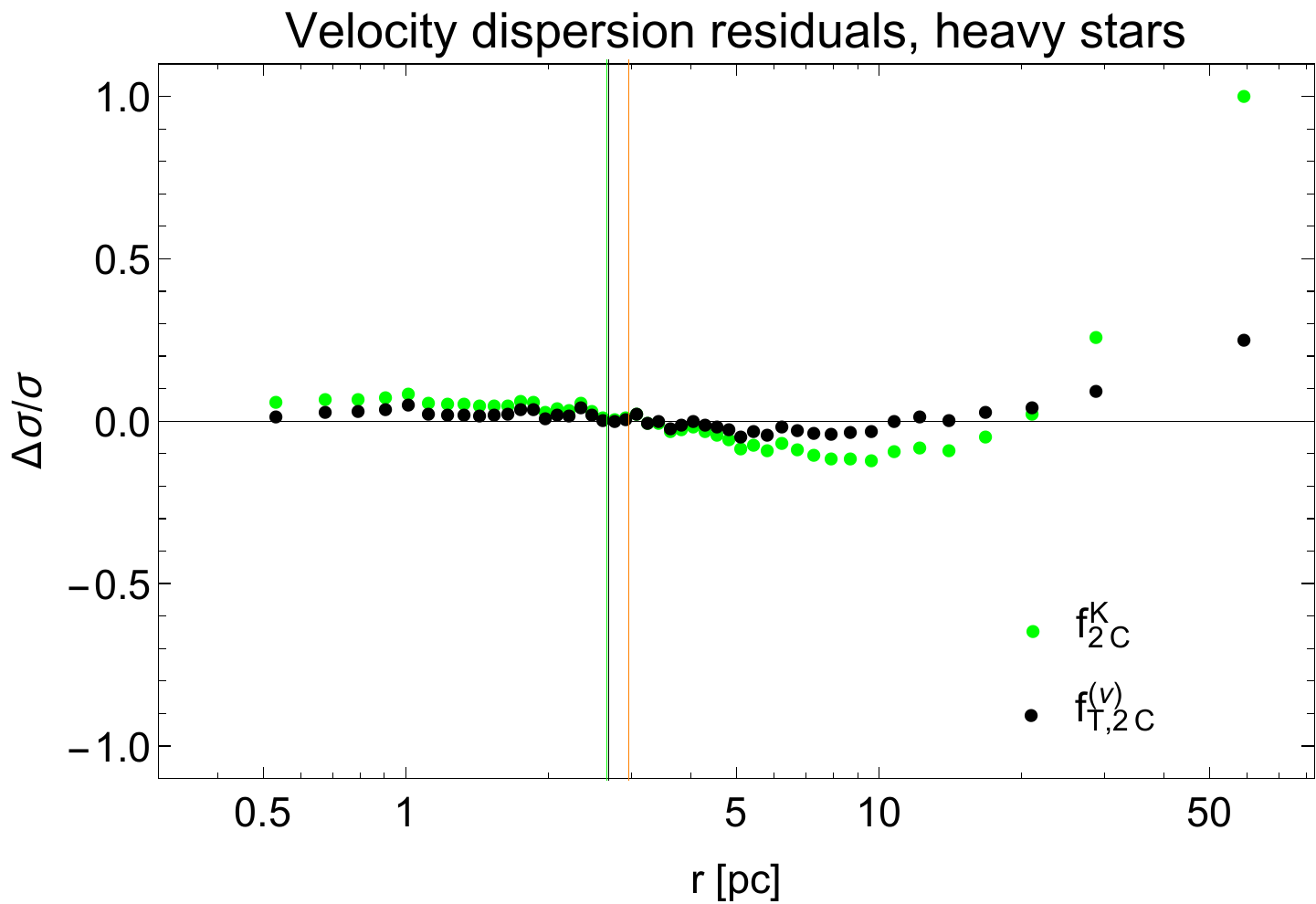} 
	\end{minipage} \hfill

		\centering
	\begin{minipage}{0.49\textwidth}
		\centering
		\includegraphics[width=0.98\textwidth]{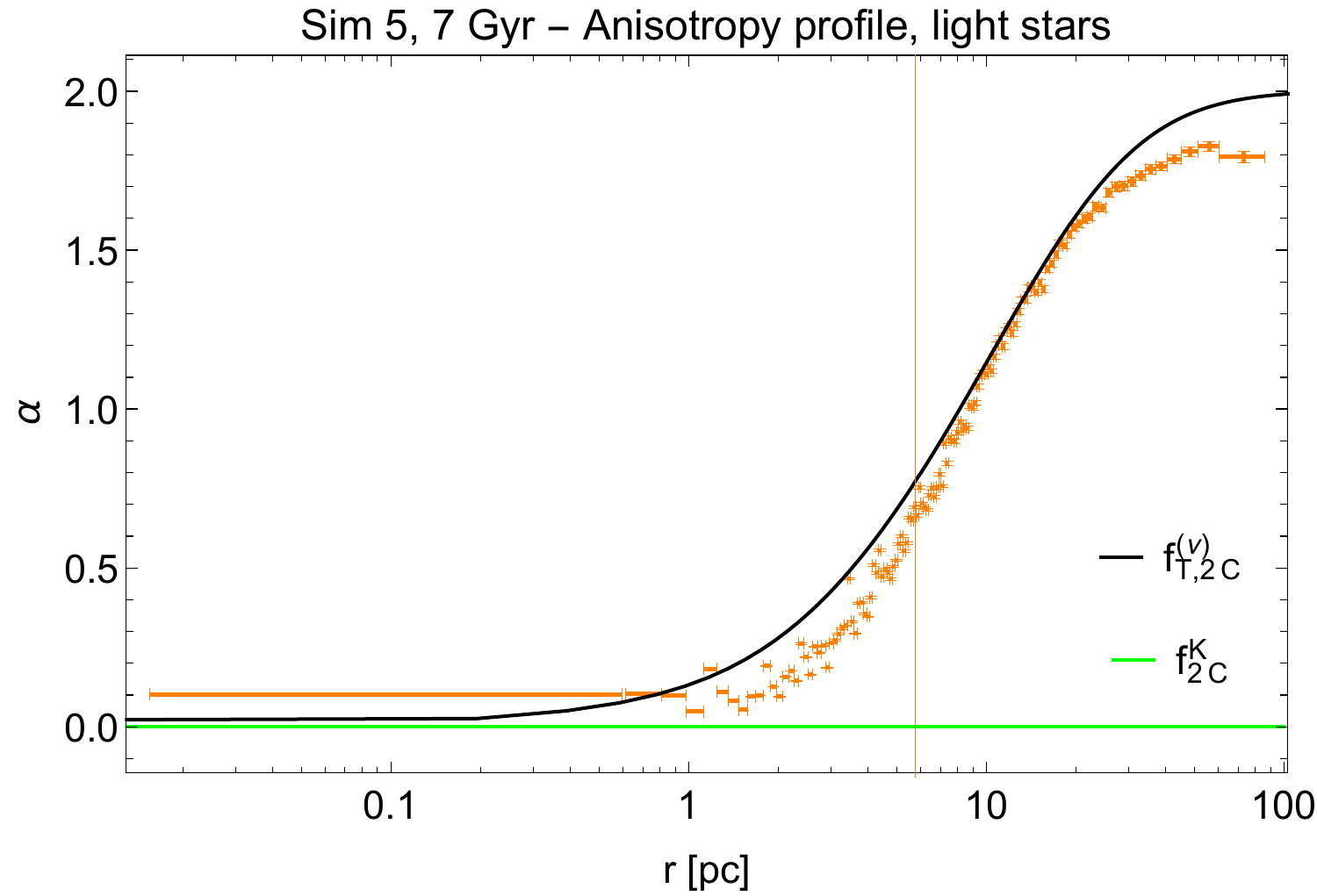}  
	\end{minipage}\hfill
	\begin{minipage}{0.49\textwidth}
		\centering
		\includegraphics[width=0.98\textwidth]{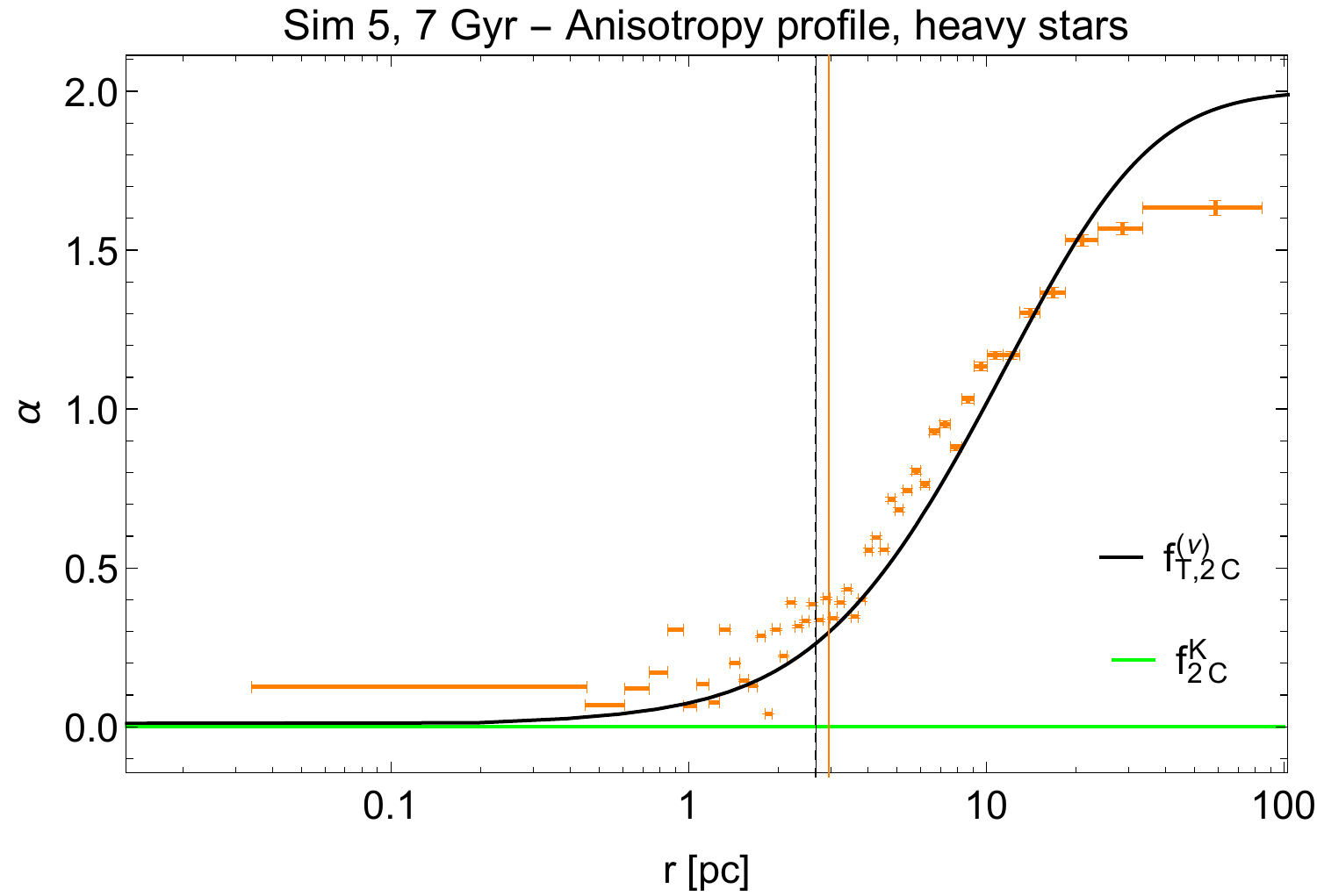} 
	\end{minipage} 
	\caption{The first four rows illustrate best-fit profiles (left column) and residuals (right column) for Sim 5 at 7 Gyr for the two-component $f_T^{(\nu)}$ models (black curves) and for the two-component King models (light, green curves). First row: light component, density profile. Second row: heavy component, density profile. Third row: light component, velocity dispersion profile. Fourth row: heavy component, velocity dispersion profile. The fifth row shows the pressure anisotropy profiles of the simulated states compared to those associated with the best-fit two-component $f_T^{(\nu)}$ models (black) and King models (light, green); the left panel refers to the light component, the right panel to the heavy component. Throughout the figure, vertical lines represent the half-mass radius of the component under consideration. }\label{fig_2c_sim5_7}

\end{figure*}

To perform the fits, we have followed a procedure very similar to that outlined by \cite[Zocchi et al. (2012)]{zocchi12}. In the present analysis, we have decided to minimize a combined chi-squared function, defined as the sum of the density and the velocity dispersion chi-squared of the two components. As an example of the comparison between models and simulated states, in Fig. \ref{fig_2c_sim5_7} we show the best-fit profiles for Sim 5 at 7 Gyr.

The two-component $f^{(\nu)}_T$ models offer a good representation of the simulated states. In fact, they give a good description of the density profiles for each of the two components of the simulated states at all radii. In addition, they are able to reproduce the central peak in the velocity dispersion profiles.

In Fig. 2, for comparison, we also show the performance of two-component models based on the King distribution function. Their density profiles present a truncation that is too sharp and central values that are too high. Their kinematic profiles exhibit large discrepancies, both in the central and in the outer regions, possibly because the underlying models are unrealistically isotropic.

Once the best-fit models are found, we compared the local degree of anisotropy, quantified by $\alpha = 2- \sigma^2_t/\sigma^2_r$, where $\sigma^2_t$ and $\sigma^2_r$ are the tangential and radial velocity dispersion (squared), of the best-fit models to that of the simulated states. The comparison is illustrated in the fifth row of Fig.~\ref{fig_2c_sim5_7}: the best-fit $f^{(\nu)}_T$ model gives a good description of the local degree of anisotropy, suggesting that the cumulative effects of collisions drive the systems toward a velocity distribution similar to that generated by violent relaxation. 

\end{document}